\documentclass[preprint,12pt]{aastex}

\shorttitle{Magnetic Fields in G240.31+0.07}
\shortauthors{Qiu et al.}
\begin{document}

\title{Submillimeter Array Observations of Magnetic Fields in G240.31+0.07: an Hourglass in a Massive Cluster-forming Core}

\author{Keping Qiu\altaffilmark{1,2}, Qizhou Zhang\altaffilmark{3}, Karl M. Menten\altaffilmark{4}, Hauyu B. Liu\altaffilmark{5}, Ya-Wen Tang\altaffilmark{5}, Josep M. Girart\altaffilmark{6}}
\email{kpqiu@nju.edu.cn}
\altaffiltext{1}{School of Astronomy and Space Science, Nanjing University, 22 Hankou Road, Nanjing 210093, China}
\altaffiltext{2}{Key Laboratory of Modern Astronomy and Astrophysics (Nanjing University), Ministry of Education, Nanjing 210093, China}
\altaffiltext{3}{Harvard-Smithsonian Center for Astrophysics, 60 Garden Street, Cambridge, MA 02138, U.S.A.}
\altaffiltext{4}{Max-Planck-Institut f\"{u}r Radioastronomie, Auf dem H\"{u}gel 69, 53121 Bonn, Germany}
\altaffiltext{5}{Academia Sinica Institute of Astronomy and Astrophysics, P. O. Box 23-141, Taipei, 106 Taiwan}
\altaffiltext{6}{Institut de Ci\`{e}ncies de l'Espai (CSIC-IEEC), Campus UAB, Facultat de Ci\`{e}ncies, C5p 2, E-08193, Bellaterra, Catalonia, Spain}

\begin{abstract}
We report the first detection of an \emph{hourglass} magnetic field aligned with a well-defined outflow-rotation system in a high-mass star-forming region. The observations were performed with Submillimeter Array toward G240.31+0.07, which harbors a massive, flattened, and fragmenting molecular cloud core and a wide-angle bipolar outflow. The polarized dust emission at 0.88~mm reveals a clear hourglass-shaped magnetic field aligned within $20^{\circ}$ of the outflow axis. Maps of high-density tracing spectral lines, e.g., H$^{13}$CO$^+$ (4--3), show that the core is rotating about its minor axis, which is also aligned with the magnetic field axis. Therefore, both the magnetic field and kinematic properties observed in this region are surprisingly consistent with the theoretical predictions of the classic paradigm of isolated low-mass star formation. The strength of the magnetic field in the plane of sky is estimated to be $\sim$1.1~mG, resulting in a mass-to-magnetic flux ratio of 1.4 times the critical value and a turbulent to ordered magnetic energy ratio of 0.4. We also find that the specific angular momentum almost linearly decreases from $r\sim0.6$~pc to 0.03~pc scales, which is most likely attributed to magnetic braking.

\end{abstract}


\section{Introduction} \label{intro}
Massive stars tend to form in clusters, which arise from the collapse and fragmentation of molecular clouds \citep{Elmegreen00}. In this process, self-gravity is certainly a necessity, since without it interstellar gas would not condense to form stars. However, a simplified theory dealing with the interplay between gravity and thermal pressure, e.g., the classic Jeans theory, fails to explain either the birth of massive stars nor the low rate of global star formation. The key question under debate is that which mechanism, turbulence or magnetic fields, plays a central role in counteracting with gravity \citep{MacLow04,Crutcher12}. In the scenario of gravo-turbulent fragmentation \citep{Klessen05,Hopkins13}, supersonic turbulence controls the evolution of molecular clouds, producing a hierarchy of density structures; as turbulence decays locally, gravitational collapse sets in to give birth to a proto-cluster. Stars near the cloud center could accrete gas at a higher rate, thus they end up being more massive. Magnetic fields are implicitly weak in this scenario, and are predicted to show a chaotic morphology due to turbulent and dynamical twisting \citep[e.g.,][]{Padoan01}. In contrast, magnetic fields are crucial in the classic theory of isolated low-mass star formation \citep{Shu87,Mouschovias99}. The theory assumes that molecular clouds are magnetically supported \citep[e.g.,][]{Ostriker01}. In collapsing cores where star formation takes place, magnetic fields are pulled by gravity into an hourglass shape; cores contract more along magnetic fields, resulting in a flattened morphology perpendicular to the field axis, and a bipolar outflow is driven along the magnetic field from an accretion disk embedded within the core \citep{Galli93,Allen03,Crutcher06}.

Observationally, hourglass magnetic fields aligned with bipolar outflows have been detected in a few low-mass objects \citep{Girart06,Stephens13}, though a recent study found that magnetic fields may be statistically misaligned with outflows \citep{Hull13}. In the high-mass regime, several high-angular-resolution observations of polarized dust emission have revealed magnetic fields with a variety of morphologies \citep[e.g.,][]{Tang09a,Tang09b,Girart13,Qiu13}. Yet a single source, G31.41+0.31, has provided the most compelling evidence for an hourglass-like magnetic field \citep{Girart09}; but the properties of gas kinematics and outflow structures are still unclear \citep{Cesaroni11,Moscadelli13}. Furthermore, there is hitherto no reported observational evidence for an hourglass magnetic field in a cluster-forming region.

Here we report Submillimeter Array\footnote{The SMA is joint project between the Smithsonian Astrophysical Observatory and the Academia Sinica Institute of Astronomy and Astrophysics and is funded by the Smithsonian Institution and the Academia Sinica.} (SMA) observations toward G240.31+0.07 (hereafter G240), which is an active high-mass star-forming region with a bolometric luminosity of $10^{4.5}~L_{\odot}$ at a distance of 5.32 kpc \citep{Choi14}. We detect a clear hourglass magnetic field threading a massive and flattened core that is powering a well-shaped bipolar outflow and fragmenting to form a proto-cluster.

\section{Observations and Data Reduction} \label{obs}
As part of an SMA polarization legacy project \citep{Zhang14}, the observations were undertaken in five tracks, each with $\sim$2--4 hours on-source integration toward the phase referencing center (R.A., Dec.)$_{\mathrm J2000}$=($07^{\mathrm h}44^{\mathrm m}51.\!^{\mathrm s}97, -24^{\circ}07{'}42.\!{''}5$). During the observations the weather conditions were excellent, with $\lesssim1$~mm precipitable water vapor, corresponding to atmospheric opacities at 225~GHz, ${\tau}_{\rm 225}$, lower than 0.06. The detailed information of the observations is presented in Table \ref{table1}. The correlator setup covers approximately 332--336 GHz in the lower sideband and 344--348 GHz in the upper sideband, with a uniform spectral resolution of 812.5 kHz ($\sim$0.7~km\,s$^{-1}$).

We performed basic data calibration, including bandpass, time dependent gain, and flux calibration, with the IDL MIR package, and output the data into MIRIAD for further processing. The intrinsic instrumental polarization was removed to a 0.1\% accuracy with the MIRIAD task GPCAL. We jointly imaged the calibrated visibilities from all five tracks to make Stokes $I$, $Q$, and $U$ maps. The continuum maps were made with a natural weighting of the visibilities to obtain an optimized sensitivity, leading to a synthesized beam of $2.\!{''}3\times1.\!{''}8$ at FWHM and an rms noise level of 0.85~mJy\,beam$^{-1}$ for Stokes $Q$, $U$ and 5~mJy\,beam$^{-1}$ for Stokes $I$; the latter is affected by the limited dynamic range. The Stokes $Q$, $U$ maps were then combined to produce maps of the linear polarization intensity, the fractional polarization, and the polarization angle (P.A.). Spectral line maps were made with a compromise of angular resolution and sensitivity, and the synthesized beam is about $1.\!{''}5\times1.\!{''}3$ and the rms noise level is $\sim45$~mJy\,beam$^{-1}$ per 0.8~km\,s$^{-1}$. No significant polarization is detected in spectral lines, so only Stokes $I$ maps are presented.

\section{Results} \label{result}
\subsection{Magnetic field morphology} \label{res_mag}
\begin{figure}
\epsscale{.95} \plotone{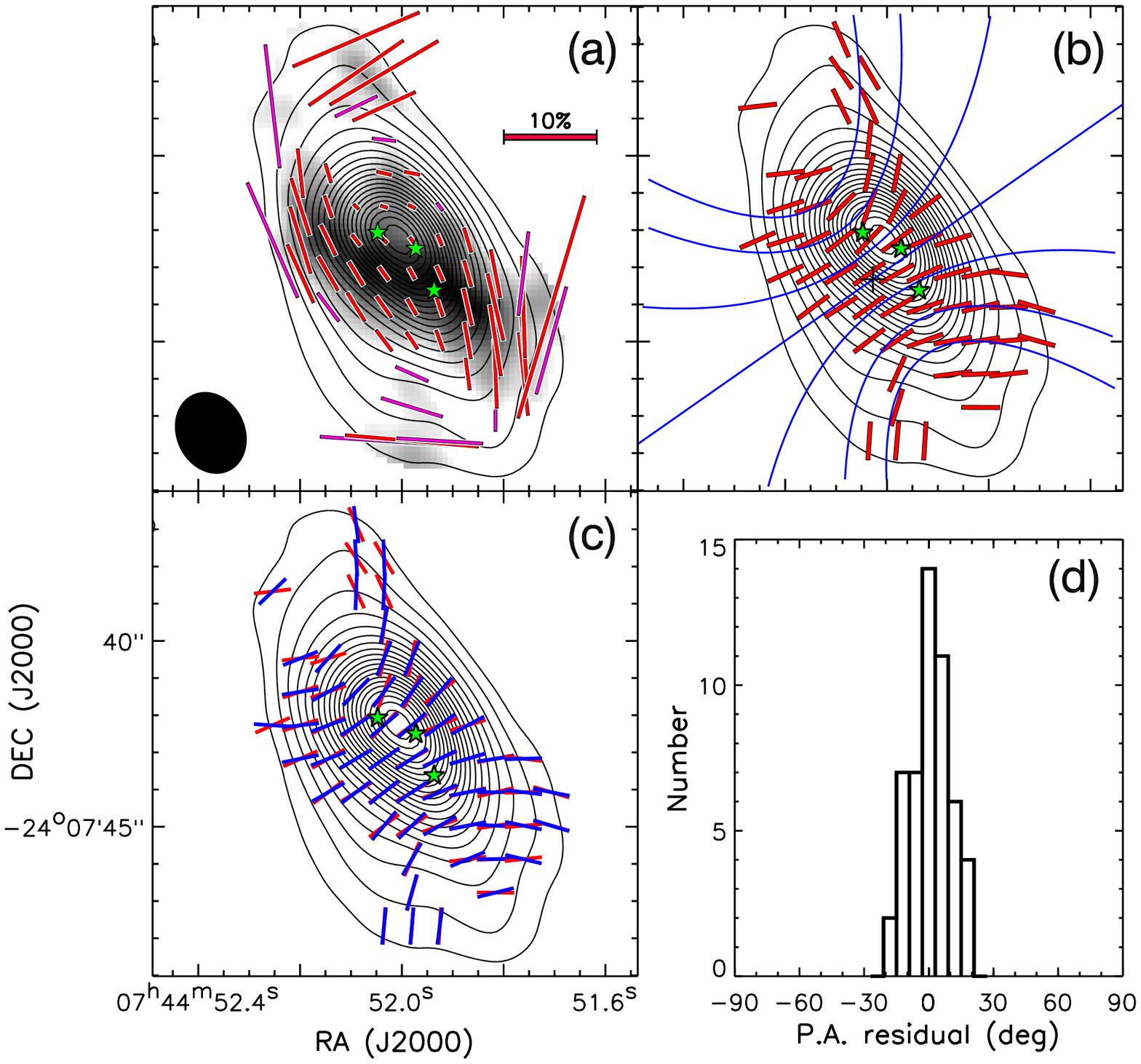}
\caption{Magnetic field morphology revealed by the polarized dust emission at 0.88~mm. (a) Contour map of the total dust emission superposed on the gray scale image the polarized emission intensity stretching from 0.0022 to 0.0073~Jy\,beam$^{-1}$. Contour levels are 0.015, 0.042, 0.078, 0.120, 0.168, 0.220, 0.278, 0.339, 0.405, 0.474, 0.547~Jy\,beam$^{-1}$. Colored bars (purple for $2.5\leq{\rm S/N}<3$ and red for ${\rm S/N}\geq3$ ) denote polarization directions, with their lengths proportional to the fractional polarization (a 10\% scale is shown in the upper right). Hereafter, a filled ellipse in the lower left shows the synthesized beam, and three star symbols mark the dense condensations (namely MM1--3 from north to south, detected in Qiu et al. 2009). (b) Contours are the same as in (a). Red bars with an arbitrary length depict magnetic field orientations, overlaid with blue curves showing a representative set of parabola derived from the best-fit model. (c) Contours are the same as in (a). Red bars are the same as those in (b). Blue bars show the modeled magnetic field orientations. (d) Histogram of the P.A. residuals for the best-fit model. \label{polmag}}
\end{figure}

We observed G240 using the SMA in its Sub-compact, Compact, and Extended configurations. In Figure \ref{polmag}, the total dust emission (Stokes $I$) reveals a flattened core, which is embedded with at least three dense condensations previously detected at 1.3~mm at a $\sim$1$''$ resolution \citep{Qiu09}. The detected polarized emission has a butterfly-like distribution (Figure \ref{polmag}a), and the inferred magnetic field exhibits a clear hourglass shape aligned with the minor axis of the core (Figure \ref{polmag}b). We fit the magnetic field morphology with an analytical model of a series of parabolic functions \citep[e.g.,][]{Girart06,Rao09}. The fitted parameters are the position angle of the magnetic field axis ${\theta}_{\rm PA}=-55\pm2^{\circ}$, the center of symmetry of the magnetic field (R.A., Dec.)$_{\mathrm J2000}$=($07^{\mathrm h}44^{\mathrm m}52.\!^{\mathrm s}03\pm0.\!^{\mathrm s}01, -24^{\circ}07{'}43.\!{''}3\pm0.\!{''}2$), and $C=0.11\pm0.01$ for the parabolic form $y=g+gCx^2$, where $x$ is the distance in units of arc second along the magnetic field axis from the center of symmetry. From Figure \ref{polmag}c, the best-fit model agrees well with the measured magnetic field. The residuals from the fit have a nearly Gaussian distribution, and yield a dispersion of $12.7\pm1.2^{\circ}$ (Figure \ref{polmag}d). Accounting for a measurement uncertainty of $7.0\pm2.5^{\circ}$, the intrinsic dispersion of the polarization angle, $\delta\phi$, is $10.6\pm2.2^{\circ}$.

\subsection{Outflow and rotation} \label{res_kin}
G240 harbors a well-shaped wide-angle outflow previously observed in CO \citep{Qiu09}, and the position angle of the outflow axis is about $-36^{\circ}$. So the bipolar outflow is roughly aligned with (within $20^{\circ}$) the magnetic field axis (Figure \ref{kinematics}a). It is already clear that we are observing an hourglass magnetic field in a massive cluster-forming core. We further investigate the kinematics of the dense gas by examining the simultaneously observed spectral lines. Figures \ref{kinematics}b and \ref{kinematics}c show the first moment map and the position-velocity (PV) diagram of the H$^{13}$CO$^+$ (4--3) emission: there is a clear velocity gradient along the major axis of the core. The velocity gradient is also detected in other high-density tracing spectral lines, and provides strong evidence for a rotating motion of the core. We perform a linear fit to the peak positions of the H$^{13}$CO$^+$ emission at each velocity channel, and derive a position angle of $-49.5\pm0.3^{\circ}$ for the rotation axis, which is only $6^{\circ}$ offset from the magnetic field axis.

\section{Discussion}
\subsection{The strength of the magnetic field}
Using the Chandrasekhar-Fermi (CF) method \citep{Chandra53}, the strength of the magnetic field in the plane of sky, $B_{\rm pos}$, can be estimated based on the P.A. dispersion, following $B_{\rm pos}=Q({\delta}V_{\rm los}/\delta\phi)(4\pi\rho)^{1/2}$, where $Q$ is a parameter related to the cloud structure and is adopted to be 0.5 \citep{Ostriker01}, ${\delta}V_{\rm los}$ is the turbulent velocity dispersion along the line of sight, and $\rho$ is the averaged mass density. In G240, $\delta\phi$ is inferred to be $10.6^{\circ}$ (Section \ref{res_mag}), and ${\delta}V_{\rm los}$ reaches 1.2~km\,s$^{-1}$ from the H$^{13}$CO$^+$ (4--3) data. To estimate the mass density, we measure the total flux of the dust emission to be 1.4~Jy over an area of 60 square arc\,sec, where the emission is detected with S/N ratios above 3. With a dust temperature of 47~K \citep{Hunter97}, a dust emissivity index of 1.5 \citep{Chen07}, and a canonical gas-to-dust mass ratio of 100, the core mass amounts to 95~$M_{\odot}$, leading to an averaged column density, $N({\rm H_2})$, of $1.5\times10^{23}$~cm$^{-2}$ and a volume density, $n({\rm H_2})$, of $2.7\times10^5$~cm$^{-3}$. Consequently, we deduce $B_{\rm pos}\sim1.1$~mG. Heitsch et al. (2001) show that the CF method is reliable to a factor of 2 for magnetic field strengths typical of molecular clouds, as long as the angular structure in the field is well resolved; limited angular resolutions lead to an overestimation of the field strength. Considering the presence of a velocity gradient across the core, the turbulent velocity dispersion inferred from the H$^{13}$CO$^+$ emission could be overestimated, resulting in an overestimation of the field strength. On the other hand, the plane-of-sky component of the field strength represents a lower limit to the total field strength. Thus, the CF method could still provide a reasonable estimate of the field strength, but the reliability needs to be verified.

Since the magnetic field has been pulled into an hourglass shape by the gravitational collapse, we can gauge the magnetic field strength from the observed curvature of the field lines. Considering the gravity and magnetic tension force acting on a field line, the magnetic field strength can be estimated following an equation derived by Schleuning (1998): $$(\frac{B}{1\,{\rm mG}})^2=(\frac{R}{0.5\,{\rm pc}})(\frac{D}{0.1\,{\rm pc}})^{-2}(\frac{M}{100\,M_{\odot}})(\frac{n({\rm H_2})}{10^5\,{\rm cm}^{-3}}),$$ where $R$ is the radius of the curvature, $D$ is the distance from the field line to the symmetry center, $M$ is the mass, which we adopt a value of 125~$M_{\odot}$ including the gas mass and the mass of the forming massive stars \citep[$\sim$30~$M_{\odot}$,][]{Trinidad11} deeply embedded within the core. From the farthest southwestern modeled line in Figure \ref{polmag}b, which has $R\sim0.047$~pc and $D\sim0.064$~pc, we obtain a field strength of 0.9~mG, which is in good agreement with the above estimate using the CF method.

\subsection{The significance of the magnetic field}
Assuming that the overall direction of the magnetic field in G240 is close to the plane of sky and $B_{\rm pos}$ approximates the total field strength, $B_{\rm tot}$, we can assess the importance of the magnetic field compared to gravity, turbulence, and rotation. The key parameter that determines whether a magnetized core would gravitationally collapse is the mass-to-magnetic flux ratio. If the ratio is greater than the critical value, $1/(2\pi\sqrt{G})$, where $G$ is the gravitational constant, the core is unstable and would collapse under its self-gravity (Nakano \& Nakamura 1978). We find the ratio in G240 is about 1.4 times the critical value, taking into account both the gas mass and the stellar mass. The core is thus slightly supercritical, and the gravity has started to overcome the magnetic force to initiate the collapse. The turbulent to ordered magnetic energy ratio, $\beta_{\rm turb}$, can be estimated from $3({\delta}V_{\rm los}/V_{\rm A})^2$, where $V_{\rm A}=B_{\rm tot}/\sqrt{4\pi\rho}$ is the Alfv\'{e}n velocity. With $B_{\rm tot}\,{\approx}\,Q({\delta}V_{\rm los}/\delta\phi)(4\pi\rho)^{1/2}$, we have $\beta_{\rm turb}\,{\approx}\,3(\delta\phi/Q)^2=0.4$, indicating that the ordered magnetic energy is dominating over the turbulent energy.

\begin{figure}
\epsscale{1.} \plotone{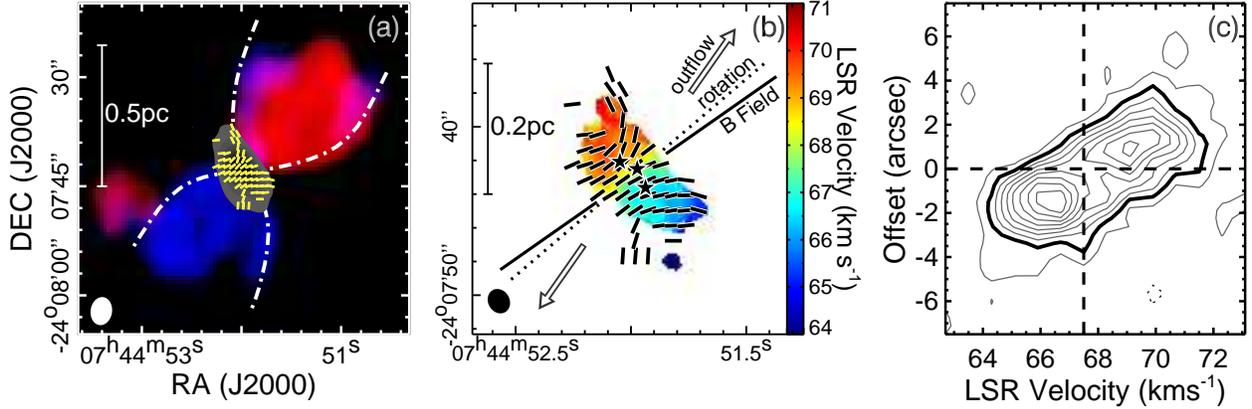}
\caption{(a) Colored image of the bipolar outflow observed in CO (2--1) \citep{Qiu09}, overlaid with the gray scale showing the 0.88~mm dust emission and yellow bars indicating the inferred magnetic field. (b) Colored image of the first moment (intensity weighted velocity) map of the H$^{13}$CO$^+$ (4--3) emission, overlaid with black bars depicting the magnetic field. Two arrows mark the orientation of the CO outflow; dotted and solid lines indicate the axes of the rotation and magnetic field, respectively. (c) Contour map of the PV diagram (cut at a position angle of $41^{\circ}$, approximately along the major axis of the core) of the H$^{13}$CO$^+$ (4--3) emission, with contour levels of 10\% to 90\%, in steps of 10\%, of the peak intensity (1.32~Jy\,beam$^{-1}$); a thick contour shows the 20\% level; a vertical dashed line indicates the systemic velocity of 67.5~km\,s$^{-1}$ \citep{Qiu09}; the PV cut is centered at the geometric center of the core, which is approximately at (R.A., Dec.)$_{\mathrm J2000}$=($07^{\mathrm h}44^{\mathrm m}52.\!^{\mathrm s}0, -24^{\circ}07{'}42.\!{''}7$) (marked by a horizontal dashed line). \label{kinematics}}
\end{figure}

\begin{figure}
\epsscale{.9} \plotone{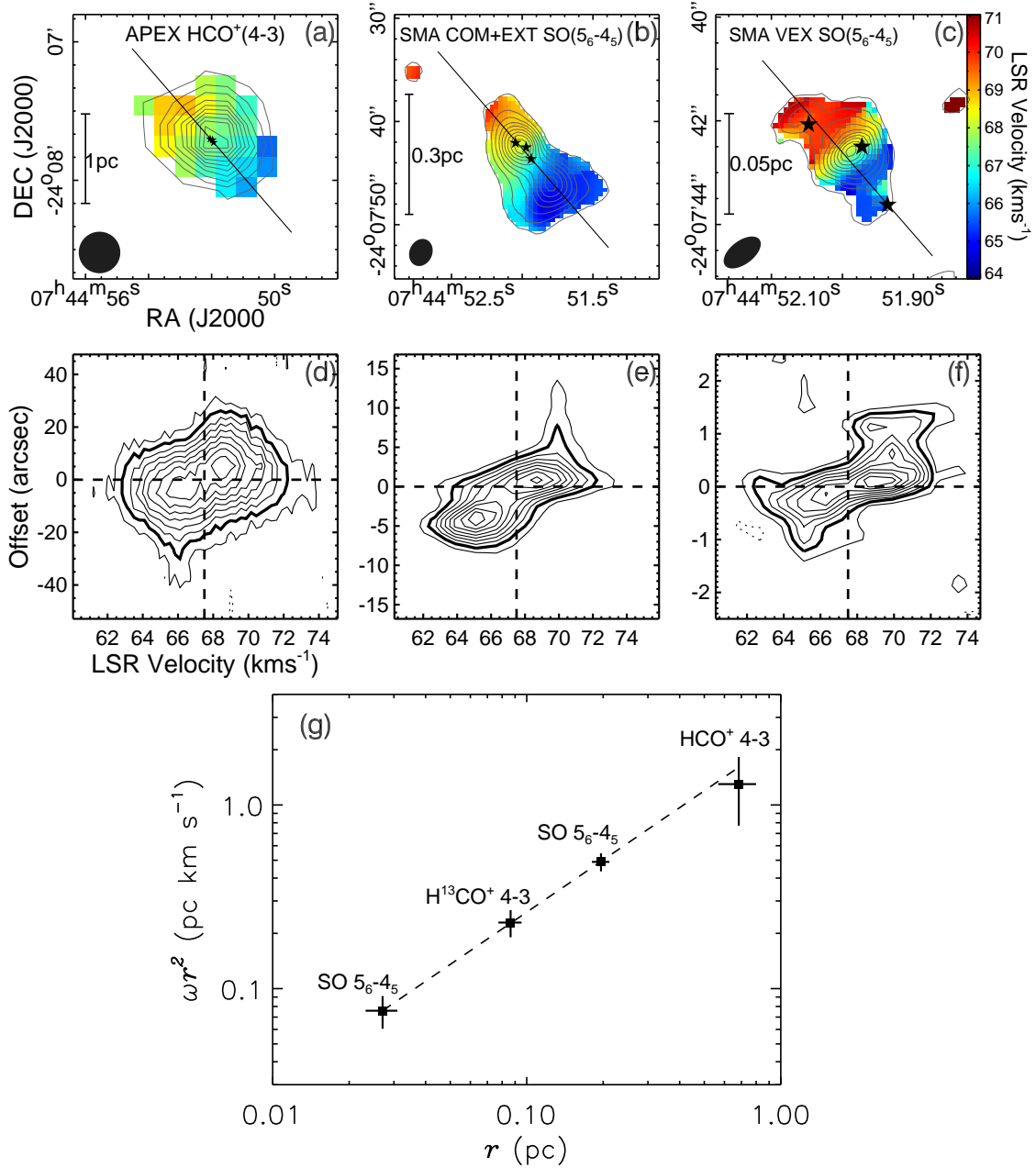}
\caption{(a--c) Colored images of the first moment maps overlaid with contours of the zeroth moment maps for the APEX HCO$^+$ (4--3), SMA COM+EXT SO ($5_6$--$4_5$), and SMA VEX SO ($5_6$--$4_5$) data; contour levels are 5\% to 95\%, in steps of 10\%, of the peak intensities for APEX HCO$^+$ (4--3) and SMA VEX SO ($5_6$--$4_5$), and 10\% to 90\%, in steps of 10\%; of the peak intensity for SMA COM+EXT SO ($5_6$--$4_5$); a solid line in each panel shows a cut at a position angle of $41^{\circ}$ for the PV diagrams shown in the following panels. (d--f) Contour maps of the PV diagrams of the above three data sets, with contour levels of 10\% to 90\%, in steps of 10\%, of the peak intensities (3.1~K in $T_{\rm A}^{\ast}$ for APEX HCO$^+$, 3.3~Jy\,beam$^{-1}$ for SMA COM+EXT SO, and 0.45~Jy\,beam$^{-1}$ for SMA VEX SO); a thick contour in each panel highlights the 20\% level. (g) Measured specific angular momenta as a function of radii; a dashed line shows the power-law fit: ${\omega}r^2=10^{0.35\pm0.10}r^{0.95\pm0.07}$. \label{angmom}}
\end{figure}

The core is observed to be rotating (Section \ref{res_kin}). For a collapsing and rotating molecular cloud core threaded by a strong magnetic field, magnetic braking is expected to channel out part of the angular momentum of the infalling matter and cause the decrease in the specific angular momentum. To check this effect, we compare the H$^{13}$CO$^+$ data from this work with our new observations made with the SMA and the Atacama Pathfinder EXperiment (APEX). In Figures \ref{angmom}a--f, we show the first moment maps and PV diagrams of SO ($5_6$--$4_5$) and HCO$^+$ (4--3). The SO maps were constructed from the SMA Very-Extended (VEX) observations and from the combined SMA Compact and Extended (COM+EXT) observations, resulting in angular resolutions of $0.83''\times0.45''$ and $2.7''\times2.2''$, respectively. The APEX HCO$^+$ data have an angular resolution of $18''$. A velocity gradient consistent with what is shown in Figure \ref{kinematics} is seen on all the maps. We measure the rotation velocities by applying the terminal velocity method \citep{Sofue01}, where the terminal velocity, $v_t$, is determined from the PV diagram by a velocity at which the intensity equals to 20\% of the peak intensity \citep[e.g.,][]{Liu13}. To mitigate the optical depth effect and the confusion from the gas in outer layers along the line of sight, we only measure the outermost half beam at FWHM of each PV diagram. With all the data, including the H$^{13}$CO$^+$ observations shown in Figure \ref{kinematics}, we obtain $v_t$ on $r\sim0.03$--0.6~pc, where the radius, $r$, is the distance from the center of the core along the PV cut, allowing us to examine how the specific angular momentum, ${\omega}r^2$, where $\omega$ ($=v_t/r$) is the angular velocity, changes with the radius. In Figure \ref{angmom}, ${\omega}r^2$ does decrease toward the center. A power-law fit to the data yields a $r^{0.95\pm0.07}$ dependence of the angular momentum on the radius, which is very likely attributed to magnetic braking.

In a numerical study, magnetic braking acts to align the rotation axis and magnetic field, while the rotation causes the magnetic field to incline through dynamo action \citep{Machida06}. When the ratio of the angular velocity to magnetic field strength, $\omega/B$, is smaller than a characteristic value of $0.39\sqrt{G}/c_s$, the former effect dominates the latter. In G240, the rotation axis is observed to be closely aligned with the magnetic field direction (Figure \ref{kinematics}b), suggesting that magnetic braking is more important. Indeed, with $\omega\sim2.6\times10^{-5}$~year$^{-1}$ from the H$^{13}$CO$^+$ data and $c_s\sim0.41$~km\,s$^{-1}$, $\omega/B$ is $2.4\times10^{-8}$~year$^{-1}\,\mu{\rm G}^{-1}$, smaller than the characteristics value of $7.8\times10^{-8}$~year$^{-1}\,\mu{\rm G}^{-1}$. Hence, magnetic braking, if it is at work as manifested by the decrease of the specific angular momentum toward the center \citep[e.g.,][]{Girart09}, should be dominating over the rotation in shaping the dynamics of the core.

\begin{figure}
\epsscale{.6} \plotone{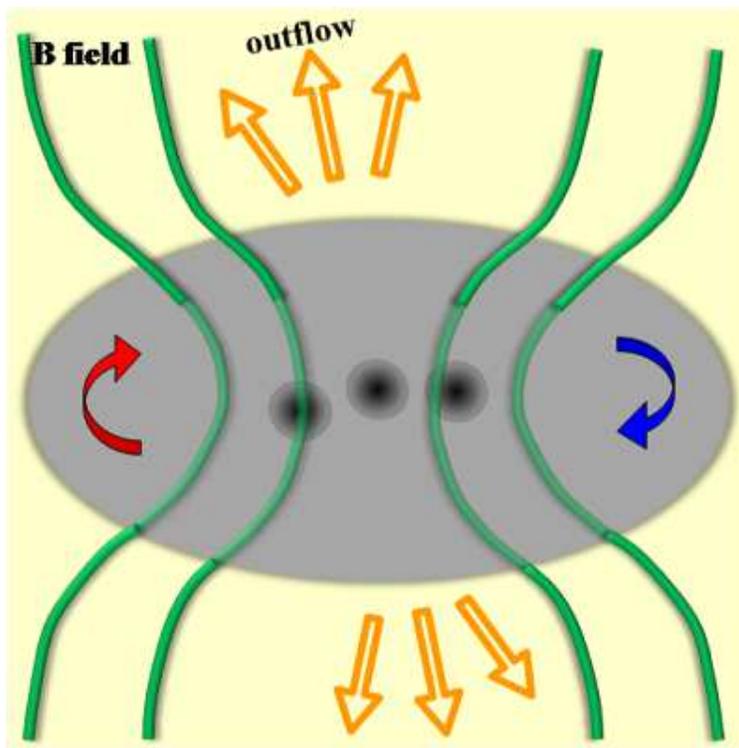}
\caption{Schematic view of the magnetic field and kinematics in G240. The core (gray-filled ellipse) is flattened, rotating (blue and red arrows indicate the sense of the rotation), and fragmenting (evidenced by the three condensations \citep{Qiu09} shown in dark gray dots), and is threaded by an hourglass magnetic field (B field) largely aligned with a bipolar outflow; the outflow is likely driven from the central condensation \citep{Qiu09}, but see Trinidad (2011) for an alternative interpretation. \label{cartoon}}
\end{figure}

We summarize the observed magnetic field and kinematic properties in G240 in Figure \ref{cartoon}. The picture is surprisingly consistent with the theoretical predictions of the classic paradigm of isolated low-mass star formation \citep{Crutcher06,Goncalves08,Frau11}, except that G240 is a luminous and massive star-forming core and shows a higher degree of fragmentation. It is the first time that a clear hourglass magnetic field aligned with a well-defined outflow-rotation system is detected in the high-mass regime. Our observations provide strong evidence that massive star and cluster formation in G240 is proceeding in a way essentially similar to the formation of Sun-like stars. We also find that the magnetic field dominates over turbulence in shaping the dynamics of the massive core. The nearly linear decrease of the specific angular momentum toward the center suggests that magnetic braking is at work and overcomes the centrifugal force to align the rotation axis.

\acknowledgments We thank the SMA staff for their supports which make this study possible. Part of this work was conducted when KQ was doing a postdoc at Max-Planck-Institut f\"{u}r Radioastronomie. KQ is partially supported by the 985 project of Nanjing University. QZ is partially supported by the NSFC grant 11328301. JMG is supported by the Spanish MINECO AYA2011-30228-C03-02, and Catalan AGAUR 2009SGR1172 grants.

\begin{deluxetable}{llllll}
\tablecaption{List of Observational Parameters \label{table1}}
\tablehead{
\colhead{Date of} & \colhead{Array} & \colhead{${\tau}_{\rm 225}$} & \colhead{Bandpass} & \colhead{Gain} & \colhead{Flux} \\
\colhead{Observations\tablenotemark{a}} & \colhead{Configuration\tablenotemark{b}} & \colhead{} & \colhead{Calibrator\tablenotemark{c}} & \colhead{Calibrator} & \colhead{Calibrator}
}
\startdata
2011 Oct. 17 (4.0) & Compact (7) & 0.03 & 3C84  & J0730-116 & Callisto \\
2011 Dec. 10 (3.5) & Compact (8) & 0.05 & 3C84  & J0730-116 & Titan    \\
2012 Jan. 06 (3.5) & Subcompact (6) & 0.06 & 3C279 & J0730-116 & Canymede \\
2012 Feb. 02 (3.5) & Extended (7) & 0.05 & 3C279 & J0730-116 & Callisto \\
2012 Mar. 27 (2.5) & Extended (7) & 0.08 & 3C279 & J0730-116 & 3C279    \\
\enddata
\tablenotetext{a}{The number in the parentheses indicates the on-source integration time in units of hours for each observation.}
\tablenotetext{b}{The number in the parentheses denotes the number of available antennas during each observation.}
\tablenotetext{c}{Also used for instrumental polarization (leakage) calibrations.}
\end{deluxetable}

\end{document}